\documentclass[a4paper]{article}
\usepackage{INTERSPEECH2020}
\usepackage{amsmath,graphicx}
\usepackage{hyperref}
\usepackage{siunitx}
\usepackage{xspace}
\usepackage{tikz}
\usepackage{pgfplots}
\usepackage{booktabs}
\usepackage{multirow}
\usepackage{csvsimple}
\usepackage{flushend}

\usepgfplotslibrary{statistics}
\pgfplotsset{compat=newest}

\usepackage[colorinlistoftodos,prependcaption,textsize=tiny]{todonotes}

\def\etal{et al.\xspace}
\def\wrt{w.r.t.\xspace}
\def\dSTOI{$\Delta\text{STOI}$\xspace}
\newcommand{\Fig}[1] {Fig.~\ref{#1}}
\newcommand{\Tab}[1] {Tab.~\ref{#1}}
\newcommand{\Sec}[1] {Sec.~\ref{#1}}

\DeclareSIUnit\k{k}
\DeclareSIUnit\M{M}
\DeclareSIUnit\FLOP{FLOP}
\DeclareSIUnit\FLOPs{FLOPs}
\DeclareSIUnit\FLOPS{FLOPS}
\DeclareSIUnit\kFLOPs{\kilo\FLOPs}
\DeclareSIUnit\kFLOPS{\kilo\FLOPS}
\DeclareSIUnit\MFLOPs{\mega\FLOPs}
\DeclareSIUnit\MFLOPS{\mega\FLOPS}

\definecolor{col1}{RGB}{211,47,47}
\definecolor{col2}{RGB}{123,31,162}
\definecolor{col3}{RGB}{0,151,167}
\definecolor{col4}{RGB}{46,125,50}

\makeatletter
\pgfplotsset{
  grid style={dotted, gray},
  boxplot/lower notch/.initial=\pgfplotsboxplotvalue{median},
  boxplot/upper notch/.initial=\pgfplotsboxplotvalue{median},
  boxplot/notch width/.initial=0.5,
  boxplot/draw/box/.code={%
    \draw[/pgfplots/boxplot/every box/.try]
      (boxplot box cs:\pgfplotsboxplotvalue{lower quartile},0)
      -- (boxplot box cs:\pgfplotsboxplotvalue{lower notch},0)
      -- (boxplot box cs:\pgfplotsboxplotvalue{median},0.5-\pgfplotsboxplotvalue{notch width}/2)
      -- (boxplot box cs:\pgfplotsboxplotvalue{upper notch},0)
      -- (boxplot box cs:\pgfplotsboxplotvalue{upper quartile},0)
      -- (boxplot box cs:\pgfplotsboxplotvalue{upper quartile},1)
      -- (boxplot box cs:\pgfplotsboxplotvalue{upper notch},1)
      -- (boxplot box cs:\pgfplotsboxplotvalue{median},0.5+\pgfplotsboxplotvalue{notch width}/2)
      -- (boxplot box cs:\pgfplotsboxplotvalue{lower notch},1)
      -- (boxplot box cs:\pgfplotsboxplotvalue{lower quartile},1)
      -- cycle
    ;
  },
  boxplot/draw/median/.code={%
    \draw[/pgfplots/boxplot/every median/.try]
        (boxplot box cs:\pgfplotsboxplotvalue{median},0.5-\pgfplotsboxplotvalue{notch width}/2)
        --
        (boxplot box cs:\pgfplotsboxplotvalue{median},0.5+\pgfplotsboxplotvalue{notch width}/2)
    ;
  }
}

\title{Lightweight Online Noise Reduction on Embedded Devices using Hierarchical Recurrent Neural Networks}
\name{H. Schr\"oter$^{1}$, T. Rosenkranz$^{2}$, A. N. Escalante-B$^{2}$, P. Zobel$^{1}$, A.
Maier$^{1}$}
\address{
  $^1$ Friedrich-Alexander-Universit\"at Erlangen-N\"urnberg, Pattern Recognition Lab\\
  $^2$ Sivantos GmbH, Research and Development, Erlangen, Germany
}
\email{hendrik.m.schroeter@fau.de}

\begin{document}
\maketitle
\begin{abstract}
Deep-learning based noise reduction algorithms have proven their success especially for non-stationary noises, which makes it desirable to also use them for embedded devices like hearing aids (HAs).
This, however, is currently not possible with state-of-the-art methods.
They either require a lot of parameters and computational power and thus are only feasible using modern CPUs.
Or they are not suitable for online processing, which requires constraints like low-latency by the filter bank and the algorithm itself.

In this work, we propose a mask-based noise reduction approach.
Using hierarchical recurrent neural networks, we are able to drastically reduce the number of neurons per layer while including temporal context via hierarchical connections.
This allows us to optimize our model towards a minimum number of parameters and floating-point operations (\si{FLOPs}), while preserving noise reduction quality compared to previous work.
Our smallest network contains only \SI{5}{\k} parameters, which makes this algorithm applicable on embedded devices.
We evaluate our model on a mixture of EUROM and a real-world noise database and report objective metrics on unseen noise.

\end{abstract}

\section{Introduction}
\label{sec:intro}

Noise reduction (NR) aims at reducing unwanted environmental noise, like street noise, and enhances a superimposed speech signal.
NR is an important feature of modern hearing aids or hearing assistance devices.
Recent contributions to deep-learning based monaural speech enhancement \cite{erdogan2015phase, pascual2017segan, aubreville2018deep, le2019phasebook, schroeter2020clcnet} result in a huge improvement over conventional noise suppression approaches \cite{ephraim1984speech, martin2001noise}.
This makes it desirable to incorporate these approaches into HAs.
However, these algorithms employing deep neural networks have great demand on both memory and computational power.
Furthermore, many algorithms process the noisy signal in an offline fashion \cite{lu2013speech, pascual2017segan, williamson2018monaural,wang2018end,zhao2018convolutional} or introduce large delays, which is not viable on HAs.
According to Jeremy \etal \cite{agnew2000just}, the maximum delay of what is typically acceptable is \SI{10}{\ms}.
Having an open acoustic coupling, a greater delay introduces annoying comb filter effects due to the superposition of the processed and direct signal.

The approaches that are close to our real-time and online processing constraints were proposed by Valin \cite{valin2018rnnoise} and Aubreville \cite{aubreville2018deep}.
Valin \etal \cite{valin2018rnnoise} uses an RNN processing \SI{20}{\ms} windows with a \SI{50}{\percent} overlap operating at a sampling rate of \SI{48}{\kHz}.
To reduce the model complexity they used a bark like scaling, which further lowers the number of input and output units.
This resulted in a network containing \SI{88.5}{\k} parameters and about \SI{40}{\MFLOPs} per second.
While this algorithm is real-time capable on a Raspberry Pi and processes the data in an online fashion, the introduced delay is greater than \SI{20}{\ms} which is too long for our requirements.

Aubreville \etal \cite{aubreville2018deep} employed a hearing instrument-grade filter bank that introduced a combined latency of analysis and synthesis of about \SI{6}{\ms}.
Additionally, they included future context of \SI{2}{\ms} resulting in an overall latency of \SI{8}{ms}.
However, they predicted Wiener gains using a fully connected network containing about \SI{28.6}{\M} parameters, resulting in about \SI{57.3}{GFLOPs} per second for the algorithm only, not including filter bank computations.

In this work, we take low latency requirements ($\le$ \SI{10}{ms}) into account and furthermore focus on a parameter and \si{\FLOPs} reduction.
To achieve our goals, we employ a uniform polyphase filter bank with a low spectral resolution.
We process our data in an about \SI{6}{\ms} frame basis with a \SI{1}{\ms} hop (\Sec{sec:signal_toolchain}).
While RNN cells, like gated recurrent units (GRUs) or long short-term memory (LSTM) cells, are able to capture long and short term dependencies, they require a sufficient amount of parameters and are hard to train.
To be able to reduce the number of parameters and thus hidden state of the recurrent state, we use a hierarchical structure to incorporate a short-term temporal context of $\pm\SI{1}{\ms}$.
This allows us to employ GRU cells with down to only 12 hidden units.
We report results on the EUROM database using 260 German sentences and 49 real-world noise signals recorded with hearing aid equipment in \Sec{sec:experiments}.
Furthermore, we provide a comparison with conventional approaches as well as previous work that employs the same processing toolchain.
We analyse the complexity of our models in \Sec{sec:complexity} and provide calculation basis and assumptions for the FLOP estimation.

\section{Signal Processing Toolchain}
\label{sec:signal_toolchain}

We use a standard uniform polyphase filter bank to transform the time domain signal into time/frequency (TF) domain.
Operating on \SI{24}{\kHz} sampling rate, the analysis window processes the input signal in an about \SI{6}{\ms} frame basis with an offset (hop) of \SI{1}{\ms}.
This filter bank ensures our low-latency requirements, but results in a low resolution spectral representation with \num{48} frequency bins.

\begin{figure}[tb]
    \centering
    \includegraphics[width=\linewidth, trim=0.5cm 9.5cm 0.5cm 0, clip]{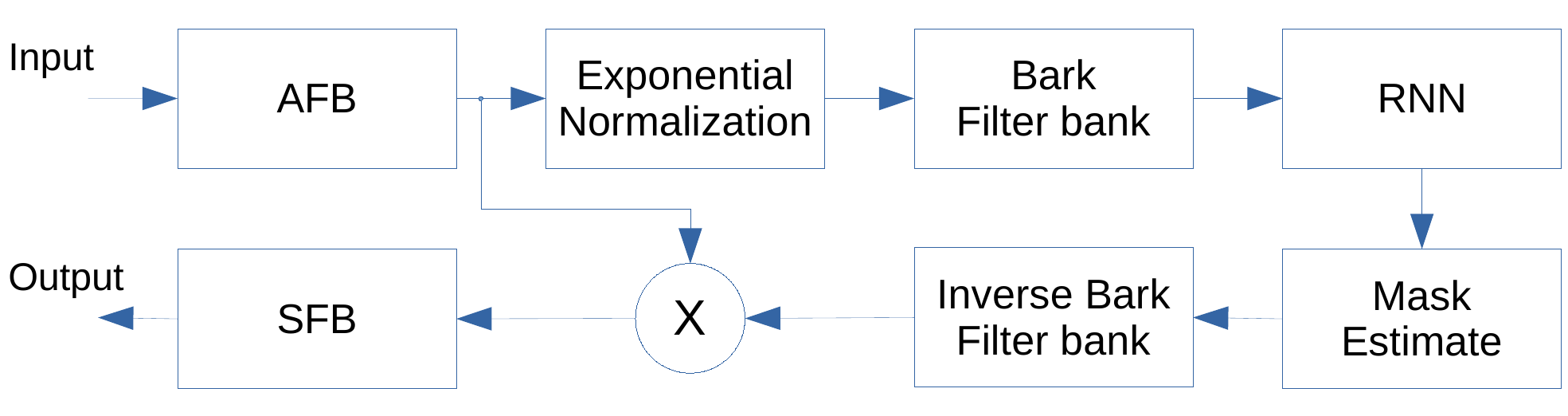}
    \caption{Processing toolchain.
        AFB/SFB depict the analysis and synthesis filter banks.}
    \label{fig:toolchain}
\end{figure}

The signal block diagram of the noise reduction system is shown in \Fig{fig:toolchain}.
We transform the complex filter bank representation into decibel scale and normalize it using exponential averaging.
The noise reduction itself is performed on a bark compressed spectral representation via a real valued mask.
The RNN is trained using the magnitude spectral approximation (MSA) loss \cite{weninger2014discriminatively}.

\subsection{Normalization}

Normalization is a crucial part of neural network training, as it tremendously helps with convergence and generalization and reduces the initialization impact.
Furthermore, since the perceived loudness by a human is scaled logarithmically, we first transform the complex filter bank representation into decibel scale and clip at \SI{-100}{\dB}.
\begin{equation}
  X_{\si{\dB}}[t, f] = 10 \cdot \log_{10} (\max(|X[t, f]|^2, 10^{-10}))\text{\ ,}
\end{equation}
where $t$ represents a time step in TF domain and $f$ a frequency band.
We then normalize the spectrogram per frequency bin to zero mean and unit variance:
\begin{equation}
  X_{\text{norm}}[t, f] = \dfrac{X_{\si{\dB}}[t, f] - \hat{\mu}[t, f]}{\sqrt{\hat{\sigma}^2[t, f]}}\text{\ ,}
\end{equation}
where $\hat{\mu}$ and $\hat{\sigma}$ are the estimated mean and variance.
That is, we can calculate mean estimates $\hat{\mu}$ and sample square estimates $\hat{s}$ via an exponential decay \cite{viikki1998recursive}:
\begin{equation}
  \hat{\mu}[t, f] = \alpha \hat{\mu}[t - 1, f] + (1 - \alpha) X_{\si{\dB}}[t, f]  \text{\ ,}
\end{equation}
\begin{equation}
  \hat{s}^2[t, f] = \alpha \hat{s}^2[t - 1, f] + (1 - \alpha) (X_{\si{\dB}}[t, f])^2  \text{\ ,}
\end{equation}
and get the variance estimate
\begin{equation}
  \hat{\sigma}^2[t, f] = \hat{s}^2[t, f] - \hat{\mu}^2[t, f]  \text{\ .}
  \label{eq:variance}
\end{equation}
Viiki \etal \cite{viikki1998recursive} suggested a normalization period $\tau$ of approximately \SI{1}{\s}.
With our sampling rate in the filter bank domain, this corresponds to $\alpha=\exp(-\Delta t/\tau)\approx0.999$.
We furthermore evaluated our model with and without variance normalization in experiment \ref{exp:normalization}.
I.e., we just assumed that the input had unit variance and skipped the square and square-root computations.
Xia \etal \cite{xia2020weighted} also used exponentially decaying normalization and compared it with global mean/variance normalization based on the training set.
They recommend the online approach based on exponential decay and a normalization period of \SI{3}{\s}.

\subsection{Bark scale}

Instead of performing the mask based noise reduction directly on the uniform filter bank representation, we further reduce the input and output dimensions using a bark like scaling of the frequency bands.
Thus, we can take advantage of the fact that human frequency perception is also on a logarithmic scale and use a coarser frequency resolution for the higher frequencies.
That is, we reduce the normalized spectrogram from 48 channels to 16 bands using rectangular bands, ensuring that the first 8 bark bands until \SI{2}{\kHz} only have 1 frequency channel and follow the bark scale for higher frequencies.
Furthermore, the network only produces a mask with 16 bins that is transformed into linear frequency scale via an inverse operation.
This allows to reduce the network size tremendously and we found that it helps a lot with convergence for very small networks.

\subsection{Hierarchical RNN}

Temporal context is essential for noise reduction algorithms.
Aubreville \etal\cite{aubreville2018deep} showed that a temporal look-back context $\gg30\si{\ms}$ is required to differentiate a fricative with little energy at low frequencies from noise only.
They furthermore showed that additionally to past context, future context also results in a better noise reduction.
However, future context always introduces a non-desirable delay, which is why they limited their future context to \SI{2}{\ms}.

\begin{figure}[tb]
    \centering
    \vspace{.1cm}
    \includegraphics[width=\linewidth]{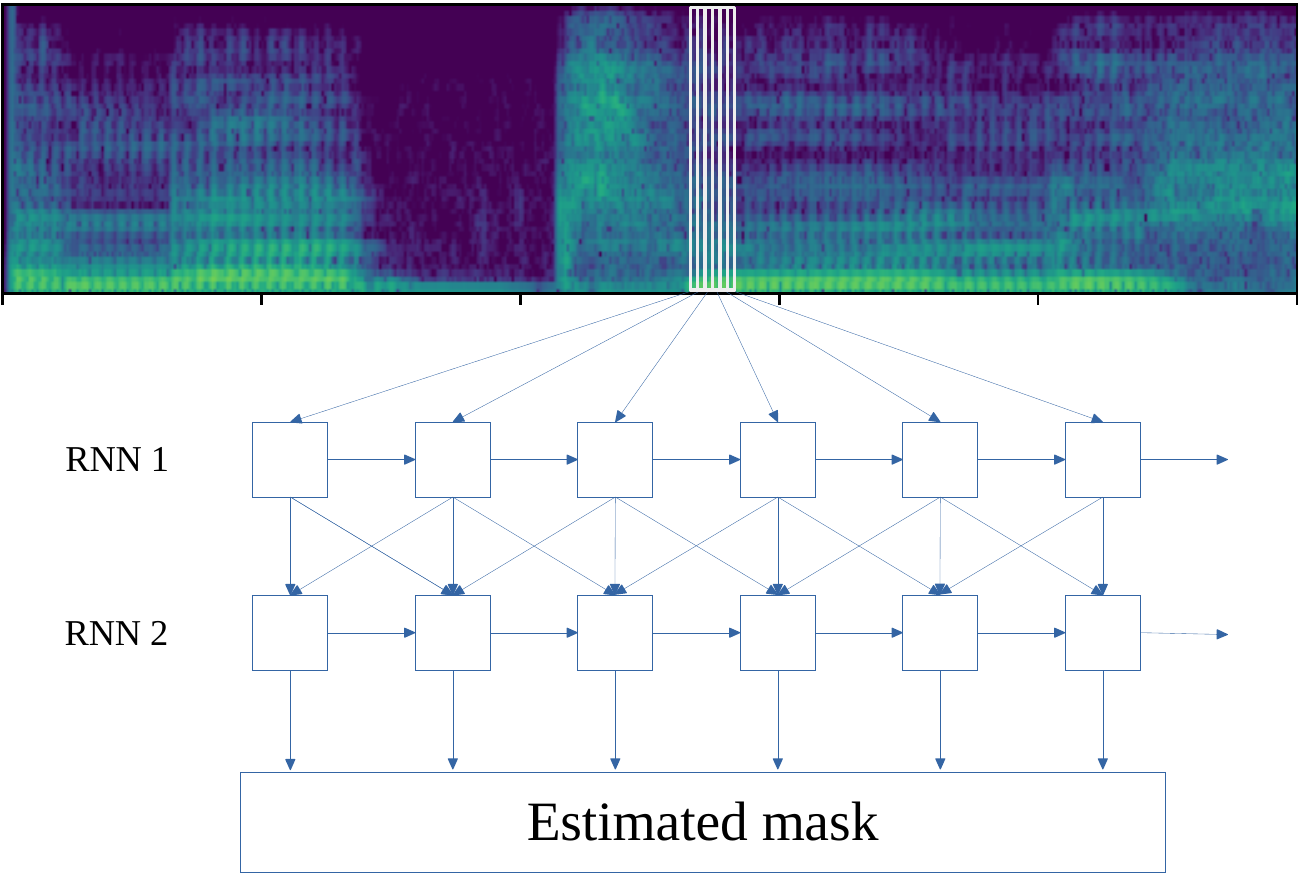}
    \vspace{.2cm}
    \caption{
      Hierarchical RNN architecture.
      The second layer RNN includes temporal context of the previous, current and next time step.
    }
    \label{fig:hrnn}
\end{figure}

Past context can obviously be incorporated using the hidden state of the RNN.
Nonetheless, the size of the hidden state, i.e., the number of neurons, limits the amount of context that can be stored.
Therefore, to allow the network to focus on the long-term temporal context such as being in a word or phoneme, we incorporate a temporal context of \SI{\pm1}{\ms}, denoted as hierarchical context (HC).
But instead of including the short-term context as input for the first RNN layer (C-RNN), we provide a context window of the first layer output as input for the second layer (HC-RNN) as shown in \Fig{fig:hrnn}.
This hierarchical short-term context inclusion is similar to fully convolutional networks that have an increasing receptive field for deeper layers due to stride $>1$.
We show in our experiments that the hierarchical structure using HC outperforms a standard RNN with early fusion.

\subsection{Network Training}

Several loss functions have been proposed in recent years.
In this work, we only focus on real-valued mask-based losses for performance and robustness reasons.
These loss functions can be broadly divided into two categories: Mask approximation (MA) and signal approximation.
For the former, common mask targets are ideal ratio mask (IRM), ideal amplitude mask (IAM) or a ``Wiener filter like'' mask (WF) \cite{erdogan2015phase}.
All of these have slightly different properties, while WF is optimal \wrt the maximum SNR, if speech and noise are uncorrelated.
Aubreville \etal successfully used a WF like mask in a hearing aids setting.
The MA loss is defined as
\begin{equation}
  L_{\text{MA}} = \sum_{t, f}(|\ |M[t, f]|- \hat{M}[t, f]\ |^2) \text{\ ,}
\end{equation}
where $\hat{M}$ is the mask estimate and $M$ the target mask, e.g. WF.
Weninger \etal \cite{weninger2014discriminatively} didn't compute the loss based on the mask, but rather forced the network to output a mask that was directly applied to the noisy signal.
The loss then was computed based on the magnitude of the resulting clean and enhanced spectrograms.
This loss is called magnitude spectrum approximation (MSA) and defined as
\begin{equation}
  L_{\text{MSA}} = \sum_{t, f}(|\ |S[t, f]|-|X[t, f]| \odot \hat{M}[t, f]\ |^2) \text{\ ,}
\end{equation}
where $\odot$ is a point-wise multiplication.
We noticed that especially for noisy conditions, MSA loss outperforms the MA loss like WF.
Additionally, we provide a comparison with a combined MA-MSA loss in experiment \ref{exp:msa_ma}.
We trained the network with gated recurrent units (GRU) layers using sequences of at least \SI{5}{\s}, a batch size of \num{20} and Adam optimizer with a learning rate of \num{0.001} using the deep learning framework PyTorch \cite{paszke2017automatic}.

\section{Experiments}
\label{sec:experiments}

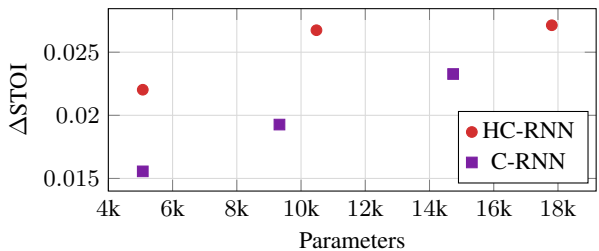
\begin{figure}[b]
  \centering
  \begin{tikzpicture}
    \begin{axis}[
      width=\columnwidth,
      height=4.0cm,
      ymin=0.014,
      xmin=4000,
      grid=both,
      grid style={solid,gray!30!white},
      xlabel={Parameters},
      ylabel={\dSTOI},
      legend style={at={(1,0)}, anchor=south east, xshift=-0.1cm, yshift=0.1cm},
      x tick label style={
        scaled x ticks = base 10:-3,
        xtick scale label code/.code={},
        xticklabel={\pgfmathprintnumber{\tick}k}
        },
      y tick label style={
        /pgf/number format/.cd,%
        scaled y ticks = false,
        precision=3,
        set decimal separator={.},
        fixed}%
      ]
      \addplot[col1, only marks] table [x=params, y=dstoi, col sep=comma] {assets/mean_dstoi_hrnn.csv};
      \addplot[col2, mark=square*, only marks] table [x=params, y=dstoi, col sep=comma] {assets/mean_dstoi_rnn_cont.csv};
      \addlegendentry{HC-RNN};
      \addlegendentry{C-RNN};
    \end{axis}
  \end{tikzpicture}
  \caption{Mean \dSTOI scores for networks of different sizes.}
  \label{fig:complexity}
\end{figure}

\noindent
In this section, we present various experiments and evaluate them using the SI-SDR metric \cite{roux2019sdr} and the difference of noisy to enhanced short-time objective intelligibility (STOI) \cite{taal2011algorithm} denoted as \dSTOI.
All networks (except from experiment \ref{exp:normalization}) use a 2 layer GRU followed by a fully connected layer with sigmoid activation.
We split our noise and speech corpus on original signal level in a train, validation (dev) and test set.
We use the same splittings as \cite{aubreville2018deep, schroeter2020clcnet}.
All results are based on the test set unless otherwise stated.

\subsection{Experiment: Variance Normalization}
\label{exp:normalization}

While many studies use zero mean and unit variance normalization \cite{aubreville2018deep, xia2020weighted} for the input, we found that normalizing to unit variance does not provide additional benefit.
\Tab{tab:exp_boxplots_norm} shows STOI improvement of a simple 3 layer LSTM network with 48 hidden units each.
There is no significant difference, which leads us to the assumption that the first network layer is able to model the input variance.
Furthermore, the network only produces an output mask that is applied to the noisy spectrogram, the mask should be independent of the noisy spectrogram scale.
Thus, for further experiments to minimize the computational effort, we only use zero mean normalization.
\begin{table}[htb]
  \caption{\dSTOI dev set results when providing input with zero mean and unit variance, and zero mean only while keeping the original variance.}
  \label{tab:exp_boxplots_norm}
  \centering
  \begin{tabular}{lr}
    \toprule
    Normalization & \dSTOI \\
    \midrule
    Zero mean \& unit variance & 0.0353 \\
    Zero mean & 0.0347 \\
    \bottomrule
  \end{tabular}
\end{table}

\subsection{Experiment: Complexity of short-term context via Hierarchical RNNs}
\label{exp:complexity-hcrnn}
\begin{table}[b]
  \caption{Network configurations with input dimension \textbf{I}, hidden dimension \textbf{H} and number of parameters \textbf{P} for the different networks in figure \ref{fig:complexity}. For all networks, the output layer is a fully connected layer with hidden dimension \num[detect-all]{16}.}
  \label{tab:network_params}
  \centering
  \robustify\bfseries
  \sisetup{
    table-number-alignment = right,
    group-minimum-digits = 4,
  }
  \begin{tabular}{l c c c c r}
    \toprule%
    & \multicolumn{2}{c}{GRU Layer 1} & \multicolumn{2}{c}{GRU Layer 2} & \multicolumn{1}{c}{Total}\\
    Context & \multicolumn{1}{c}{\# \textbf I} & \multicolumn{1}{c}{\# \textbf H} & \multicolumn{1}{c}{\# \textbf I} & {\# \textbf H} & \multicolumn{1}{c}{\# \textbf P} \\
    \midrule%
    C-RNN  & \num{48} & \num{16} & \num{16} & \num{16} & \num{5072}\text{ } \\
    C-RNN  & \num{48} & \num{24} & \num{24} & \num{24} & \num{9382}\text{ } \\
    C-RNN  & \num{48} & \num{32} & \num{32} & \num{32} & \num{14736}\text{ }\\
    HC-RNN & \num{16} & \num{16} & \num{48} & \num{16} & \num{5072}\text{ } \\
    HC-RNN & \num{16} & \num{24} & \num{72} & \num{24} & \num{10480}\text{ }\\
    HC-RNN & \num{16} & \num{32} & \num{96} & \num{32} & \num{17808}\text{ }\\
    \bottomrule%
  \end{tabular}
\end{table}
\begin{table*}[t]
  \caption{
    Metric results on the test set using our smallest HC-RNN model, RNNoise \cite{valin2018rnnoise} and FC-WF \cite{aubreville2018deep}.
    Number of network parameters in brackets.
    SI-SDR and \dSTOI are provided depending on the input SNR, the root mean squared error of the time-domain signal (RMSE) is averaged over all input SNRs. The recursive minimum tracking baseline \cite{hansler2005acoustic} is also evaluated using an attenuation limit of \SI{14}{\dB}.
  }
  \label{tab:exp_metric_results}
  \centering
  \robustify\bfseries
  \sisetup{
    table-number-alignment = center,
    table-figures-integer  = 1,
    table-figures-decimal  = 2,
    table-auto-round = true,
    detect-weight = true
  }
  \resizebox{\textwidth}{!}{
    \begin{tabular}{l@{\hskip 0.5cm}
      S S S S S@{\hskip 0.7cm}
      S[table-figures-decimal=3]
      S[table-figures-decimal=3]
      S[table-figures-decimal=3]
      S[table-figures-decimal=3]
      S[table-figures-decimal=3]@{\hskip 0.7cm}
      S[table-figures-decimal=3]
    }%
      \toprule%
      \multirow{2}{*}{Metric per SNR [\si{\dB}]} & \multicolumn{5}{c}{SI-SDR [\si{\dB}]} & \multicolumn{5}{c}{\dSTOI} & \multicolumn{1}{c}{RMSE} \\
      &
        \num{-5} & \num{0} & \num{5} & \num{10} & \num{20} &
        \num{-5} & \num{0} & \num{5} & \num{10} & \num{20} &
        \hspace{0.65cm}\text{-}%
      \\\midrule\vspace{-1em}%
      \bfseries Model &&&&&&&&&&& \\
      \begin{filecontents*}{assets/metrics-all.csv}
name,si_sdr_id,sdra,sdrb,sdrc,sdrd,sdre,dstoi_id,dstoia,dstoib,dstoic,dstoid,dstoie,rmse_id,rmse
HC-RNN$_{MSA}$ (\SI{5}{k}),si-sdr-mask_wg_context_msa_rmse12.csv,\bfseries 0.8194787264376198,5.460442520104921,9.705194633740645,\bfseries 13.80957871522659,\bfseries 21.706609759575283,dstoi-mask_wg_context_msa_rmse12.csv,\bfseries 0.025447374496322412,0.03798075889547666,0.026918055155338384,0.014583880989215313,0.0025240841966408948,rmse-mask_wg_context_msa_rmse12.csv,0.021552263182970002
HC-RNN$_{MA-MSA}$ (\SI{5}{k}),si-sdr-mask_wg_context_ma_msa_rmse21.csv,0.5591907060431998,5.178986265491217,9.300625656659786,13.432546517787836,21.677478726093586,dstoi-mask_wg_context_ma_msa_rmse21.csv,0.023424029923402347,\bfseries 0.04162803827187954,0.03556053072978289,0.024180168620286845,\bfseries 0.007500895991539344,rmse-mask_wg_context_ma_msa_rmse21.csv,0.02216246517267651
RNNoise (\SI{68}{k}) \cite{valin2018rnnoise},si-sdr-rnnoise.csv,-0.836837132817891,\bfseries 6.4562225830860624,\bfseries 10.143817378924442,12.713000049957863,15.448767885183676,dstoi-rnnoise.csv,-0.004804559195270905,0.03156579524660722,0.03233979499110809,0.02360272866029006,0.006557680188845365,rmse-rnnoise.csv,\bfseries 0.020843502830860252
FC-WF (\SI{25}{M}) \cite{aubreville2018deep},si-sdr-kai-fc-wg.csv,-1.6463735581825623,3.5883624204404794,8.443577302468789,12.843060984582959,21.405701157674624,dstoi-kai-fc-wg.csv,0.019440556677539696,\bfseries 0.04190175966092735,\bfseries 0.039249385369790564,\bfseries 0.026626576206641283,0.007219805538309792,rmse-kai-fc-wg.csv,0.024878937531625418
Baseline,si-sdr-fancy.csv,-5.68065609353093,0.9091458994244255,5.96769713018185,10.174102564652761,16.940612071599716,dstoi-fancy.csv,-0.1438635228536068,-0.0946610858425116,-0.059111389403159804,-0.04024325903409567,-0.02616838384897281,rmse-fancy.csv,0.028442296749562956
Baseline$_{\SI{14}{\dB}}$,si-sdr-fancy-14dB.csv,-4.070517424758202,1.7411643300470014,6.565517242902365,10.66555280257494,17.48038442012591,dstoi-fancy-14dB.csv,-0.005583311550510235,-0.01093461689276573,-0.015201892990332384,-0.015941025641484138,-0.015095629466649814,rmse-fancy-14dB.csv,0.027141780471757578
\end{filecontents*}
\csvreader[head to column names]{assets/metrics-all.csv}{}%
      {\\\csvcoli&\sdra&\sdrb&\sdrc&\sdrd&\sdre&\dstoia&\dstoib&\dstoic&\dstoid&\dstoie&\rmse}%
      \\\bottomrule%
    \end{tabular}
  }
\end{table*}

In this experiment, we show that networks with hierarchical context structure (HC-RNN) are favorable \wrt \dSTOI score.
While HC-RNNs only need a small input layer, the second layer incorporates the short-term context of $\pm 1$ frames and thus is larger.
We compare these with networks that incorporate the short-term context in the first layer and thus, have a larger 1st layer than 2nd layer (C-RNN).
\Fig{fig:complexity} shows \dSTOI scores of various network sizes.
All network configurations are shown in table \ref{tab:network_params}.

\noindent
We can see that hierarchical RNN outperforms the conventional RNN that includes context in the first layer on a per parameter basis.
This leads us to conclude that at least for these small numbers of parameters, it is beneficial to integrate the short-term context at later layer and thus shift the majority of parameters to deeper layers.

\subsection{Experiment: Mask Approximation vs. Signal Approximation}
\label{exp:msa_ma}

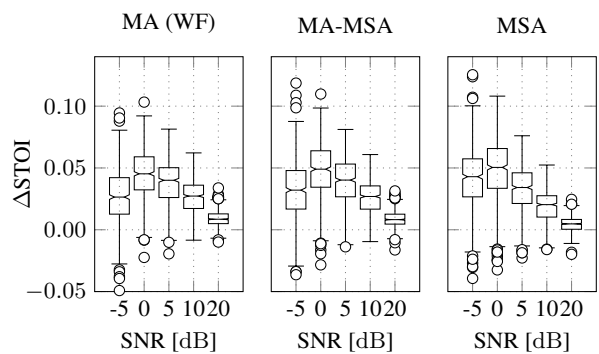
\begin{figure}[bth]
  \centering
  \resizebox{\columnwidth}{!} {
    \begin{tikzpicture}
  \begin{axis}[
    title={MA (WF)},
    xmin=0,
    xmax=6,
    xlabel={SNR [\si{\dB}]},
    xtick={1, 2, 3, 4, 5},
    xticklabels={-5, 0, 5, 10, 20},
    ymin=-0.05,
    ymax=0.14,
    ylabel={\dSTOI},
    ylabel style={
      yshift=-2ex,
    },
    y tick label style={
      /pgf/number format/.cd,
        fixed, fixed zerofill, precision=2,
      /tikz/.cd
    },
    grid=major,
    boxplot/draw direction=y,
    width=0.45\linewidth,
    height=0.6\linewidth,
    boxplotcolor/.style={color=#1,mark options={color=#1,fill=white}},
    ]
    \addplot+[boxplotcolor=black, mark=*,boxplot prepared={
        median=0.026441842317581177,
        lower quartile=0.012636363506317139,
        upper quartile=0.04214818775653839,
        lower whisker=-0.027641624212265015,
        upper whisker=0.08054053783416748,
        lower notch=0.024587015299032487,
        upper notch=0.028296669336129867}
        ]
        coordinates {
             (1.0, -0.032420068979263306) (1.0, -0.05284535884857178) (1.0, -0.03400564193725586) (1.0, -0.037997037172317505) (1.0, -0.039450377225875854) (1.0, -0.049178481101989746) (1.0, 0.09449785947799683) (1.0, 0.0878869891166687) (1.0, 0.09042680263519287) 
        };
    \addplot+[boxplotcolor=black, mark=*,boxplot prepared={
        median=0.04526051878929138,
        lower quartile=0.03228782117366791,
        upper quartile=0.05901975929737091,
        lower whisker=-0.0061980485916137695,
        upper whisker=0.09211486577987671,
        lower notch=0.04358040844878458,
        upper notch=0.046940629129798184}
        ]
        coordinates {
             (1.0, -0.008424997329711914) (1.0, -0.02239888906478882) (1.0, -0.008115410804748535) (1.0, 0.1032646894454956) 
        };
    \addplot+[boxplotcolor=black, mark=*,boxplot prepared={
        median=0.040040820837020874,
        lower quartile=0.0260746031999588,
        upper quartile=0.05021741986274719,
        lower whisker=-0.008580446243286133,
        upper whisker=0.08141458034515381,
        lower notch=0.03852343755802308,
        upper notch=0.04155820411601867}
        ]
        coordinates {
             (1.0, -0.019465744495391846) (1.0, -0.01014411449432373) 
        };
    \addplot+[boxplotcolor=black, mark=*,boxplot prepared={
        median=0.027262389659881592,
        lower quartile=0.01722124218940735,
        upper quartile=0.03606395423412323,
        lower whisker=-0.008426487445831299,
        upper whisker=0.06218236684799194,
        lower notch=0.026078119548114106,
        upper notch=0.028446659771649077}
        ]
        coordinates {
             
        };
    \addplot+[boxplotcolor=black, mark=*,boxplot prepared={
        median=0.00853988528251648,
        lower quartile=0.004964783787727356,
        upper quartile=0.01285320520401001,
        lower whisker=-0.0068176984786987305,
        upper whisker=0.02425682544708252,
        lower notch=0.008044095627069833,
        upper notch=0.009035674937963126}
        ]
        coordinates {
             (1.0, -0.00813835859298706) (1.0, -0.00995337963104248) (1.0, 0.025956153869628906) (1.0, 0.026757359504699707) (1.0, 0.025130867958068848) (1.0, 0.02656090259552002) (1.0, 0.025576770305633545) (1.0, 0.026164531707763672) (1.0, 0.02680259943008423) (1.0, 0.02858024835586548) (1.0, 0.025454342365264893) (1.0, 0.02586430311203003) (1.0, 0.026587605476379395) (1.0, 0.02611255645751953) (1.0, 0.02742713689804077) (1.0, 0.033847153186798096) (1.0, 0.025080323219299316) 
        };
  \end{axis}
\end{tikzpicture}
    \begin{tikzpicture}
  \begin{axis}[
    title={MA-MSA},
    xmin=0,
    xmax=6,
    xlabel={SNR [\si{\dB}]},
    xtick={1, 2, 3, 4, 5},
    xticklabels={-5, 0, 5, 10, 20},
    ymin=-0.05,
    ymax=0.14,
    ytick={},
    yticklabels={},
    ylabel={},
    y tick label style={
      /pgf/number format/.cd,
        fixed, fixed zerofill, precision=2,
      /tikz/.cd
    },
    grid=major,
    boxplot/draw direction=y,
    width=0.45\linewidth,
    height=0.6\linewidth,
    boxplotcolor/.style={color=#1,mark options={color=#1,fill=white}},
    ]
    \addplot+[boxplotcolor=black, mark=*,boxplot prepared={
        median=0.03205764293670654,
        lower quartile=0.016748785972595215,
        upper quartile=0.047901809215545654,
        lower whisker=-0.02937227487564087,
        upper whisker=0.0876152515411377,
        lower notch=0.03009966606850035,
        upper notch=0.03401561980491274}
        ]
        coordinates {
             (1.0, -0.05568432807922363) (1.0, -0.0338464081287384) (1.0, -0.03384026885032654) (1.0, -0.0363713800907135) (1.0, 0.11865508556365967) (1.0, 0.10878074169158936) (1.0, 0.10278278589248657) (1.0, 0.09867322444915771) 
        };
    \addplot+[boxplotcolor=black, mark=*,boxplot prepared={
        median=0.0490131676197052,
        lower quartile=0.03441859781742096,
        upper quartile=0.06389662623405457,
        lower whisker=-0.008899867534637451,
        upper whisker=0.09848731756210327,
        lower notch=0.0471604646794487,
        upper notch=0.0508658705599617}
        ]
        coordinates {
             (1.0, -0.016447067260742188) (1.0, -0.01082003116607666) (1.0, -0.02840358018875122) (1.0, -0.01942288875579834) (1.0, -0.012631893157958984) (1.0, -0.011286258697509766) (1.0, 0.10979330539703369) 
        };
    \addplot+[boxplotcolor=black, mark=*,boxplot prepared={
        median=0.04019629955291748,
        lower quartile=0.026673853397369385,
        upper quartile=0.053151264786720276,
        lower whisker=-0.01208186149597168,
        upper whisker=0.08115530014038086,
        lower notch=0.03853218629411652,
        upper notch=0.04186041281171844}
        ]
        coordinates {
             (1.0, -0.013797760009765625) 
        };
    \addplot+[boxplotcolor=black, mark=*,boxplot prepared={
        median=0.026923060417175293,
        lower quartile=0.016684547066688538,
        upper quartile=0.03549857437610626,
        lower whisker=-0.009575188159942627,
        upper whisker=0.06084263324737549,
        lower notch=0.025740593149637288,
        upper notch=0.028105527684713298}
        ]
        coordinates {
             
        };
    \addplot+[boxplotcolor=black, mark=*,boxplot prepared={
        median=0.008191436529159546,
        lower quartile=0.004571869969367981,
        upper quartile=0.012611910700798035,
        lower whisker=-0.00722736120223999,
        upper whisker=0.024569392204284668,
        lower notch=0.007686117554214219,
        upper notch=0.008696755504104873}
        ]
        coordinates {
             (1.0, -0.0163271427154541) (1.0, -0.010515213012695312) (1.0, -0.00801074504852295) (1.0, 0.025120317935943604) (1.0, 0.027823269367218018) (1.0, 0.025968074798583984) (1.0, 0.029870986938476562) (1.0, 0.0281180739402771) (1.0, 0.025676250457763672) (1.0, 0.025739431381225586) (1.0, 0.03144031763076782) 
        };
  \end{axis}
\end{tikzpicture}
    \begin{tikzpicture}
  \begin{axis}[
    title={MSA},
    xmin=0,
    xmax=6,
    xlabel={SNR [\si{\dB}]},
    xtick={1, 2, 3, 4, 5},
    xticklabels={-5, 0, 5, 10, 20},
    ymin=-0.05,
    ymax=0.14,
    ytick={},
    yticklabels={},
    ylabel={},
    y tick label style={
      /pgf/number format/.cd,
        fixed, fixed zerofill, precision=2,
      /tikz/.cd
    },
    grid=major,
    boxplot/draw direction=y,
    width=0.45\linewidth,
    height=0.6\linewidth,
    boxplotcolor/.style={color=#1,mark options={color=#1,fill=white}},
    ]
    \addplot+[boxplotcolor=black, mark=*,boxplot prepared={
        median=0.04293045401573181,
        lower quartile=0.026628226041793823,
        upper quartile=0.057379961013793945,
        lower whisker=-0.017980337142944336,
        upper whisker=0.10026997327804565,
        lower notch=0.040997698235889196,
        upper notch=0.04486320979557443}
        ]
        coordinates {
             (1.0, -0.02116018533706665) (1.0, -0.02447265386581421) (1.0, -0.027103900909423828) (1.0, -0.030495882034301758) (1.0, -0.029843568801879883) (1.0, -0.03936183452606201) (1.0, -0.023841917514801025) (1.0, 0.12378466129302979) (1.0, 0.12541872262954712) (1.0, 0.10720300674438477) (1.0, 0.10621446371078491) 
        };
    \addplot+[boxplotcolor=black, mark=*,boxplot prepared={
        median=0.05044242739677429,
        lower quartile=0.03353479504585266,
        upper quartile=0.06571991741657257,
        lower whisker=-0.013659238815307617,
        upper whisker=0.10810911655426025,
        lower notch=0.04841958278837772,
        upper notch=0.052465272005170864}
        ]
        coordinates {
             (1.0, -0.01938694715499878) (1.0, -0.024422764778137207) (1.0, -0.015620768070220947) (1.0, -0.021395385265350342) (1.0, -0.016324162483215332) (1.0, -0.018275797367095947) (1.0, -0.024346649646759033) (1.0, -0.03248018026351929) (1.0, -0.02563077211380005) (1.0, -0.018043339252471924) 
        };
    \addplot+[boxplotcolor=black, mark=*,boxplot prepared={
        median=0.03417566418647766,
        lower quartile=0.021264120936393738,
        upper quartile=0.04608041048049927,
        lower whisker=-0.013086676597595215,
        upper whisker=0.07604736089706421,
        lower notch=0.03261595293459905,
        upper notch=0.035735375438356275}
        ]
        coordinates {
             (1.0, -0.022819697856903076) (1.0, -0.016852080821990967) (1.0, -0.018623411655426025) (1.0, -0.016106784343719482) (1.0, -0.018619298934936523) 
        };
    \addplot+[boxplotcolor=black, mark=*,boxplot prepared={
        median=0.02033829689025879,
        lower quartile=0.010269656777381897,
        upper quartile=0.027537792921066284,
        lower whisker=-0.014629185199737549,
        upper whisker=0.05237776041030884,
        lower notch=0.01925298934682413,
        upper notch=0.021423604433693448}
        ]
        coordinates {
             (1.0, -0.01565760374069214) (1.0, -0.016019165515899658) 
        };
    \addplot+[boxplotcolor=black, mark=*,boxplot prepared={
        median=0.004740387201309204,
        lower quartile=0.00040359795093536377,
        upper quartile=0.008348643779754639,
        lower whisker=-0.011068165302276611,
        upper whisker=0.019563496112823486,
        lower notch=0.004241038684526441,
        upper notch=0.0052397357180919675}
        ]
        coordinates {
             (1.0, -0.018201053142547607) (1.0, -0.01985985040664673) (1.0, 0.0209273099899292) (1.0, 0.024600088596343994) (1.0, 0.020638108253479004) 
        };
  \end{axis}
\end{tikzpicture}
  }
  \caption{\dSTOI results using a hierarchical RNN with \SI[detect-all]{25}{\k} parameters trained with MSA loss, MA loss using a WF like mask and combination.}
  \label{fig:exp_boxplots_msa}
\end{figure}

Prior work reported that an MSA objective outperforms MA approaches like WF \cite{erdogan2015phase, weninger2014discriminatively}.
However, we found that this heavily depends on the input SNR.
The box plot in \Fig{fig:exp_boxplots_msa} shows that MSA outperforms MA with a WF like mask in noisy conditions (SNR $\le 0$).
For SNRs $\ge$ 5, MA and MA-MSA loss results in better \dSTOI scores.
Since MA-MSA and MSA seem to perform similarly, we evaluated both for our smallest model with \SI{5}{\k} parameters in experiment \ref{exp:objective_evaluation}.

\subsection{Experiment: Objective evaluation and comparison with prior work}
\label{exp:objective_evaluation}

We compare our smallest HC-RNN model with various state-of-the-art results.
As a baseline we chose recursive minimum tracking \cite{hansler2005acoustic} and also provide results using FC-WG \cite{aubreville2018deep} and RNNoise \cite{valin2018rnnoise} based on our test set.
%
%
As one can see in \Tab{tab:exp_metric_results}, the HC-RNN models performs similarly to prior studies, while the performance slightly drops for larger SNRs for the MSA loss.
In any case, it still outperforms conventional methods like recursive minimum tracking \cite{hansler2005acoustic}, that is only able to provide an SDR improvement for SNRs in range \numrange{0}{10}.

\section{Complexity analysis}
\label{sec:complexity}

The number of \si{\FLOPs} per second of a GRU is given by 
\begin{equation} T \cdot 6N(M+N+1) \end{equation}
where $T$ are the number of time steps, $M$ the input size and $N$ the hidden size.
In our case, we have $T=1000$ steps per second, the smallest network has \num{16} hidden units in each layer, while the input size is \num{16} and \num{48} respectively.
We count a separate operation for multiply and add.
For activation functions like tanh or sigmoid, we assume a lookup table consuming \SI{1}{\FLOP}.
Additionally with the fully connected output layer, this results in \SI{10.0}{\MFLOPS}.
The normalization requires additional \SI{0.14}{\MFLOPS}, the bark scaling \SI{48}{\kFLOPS}. The filter bank is not considered here, since it is required also for other tasks and implemented via hardware.
Compared to RNNoise which requires about \SI{40}{\MFLOPS}, we still have a considerable \si{\FLOPS} reduction of a factor \num{4}, while lowering the delay from \SI{\ge20}{\ms} to \SI{8}{\ms}.
Our model is able to run in real-time on a Raspberry Pi 3 using the non optimized Python front-end of the deep learning framework PyTorch.

\section{Conclusion}
\label{sec:conclusion}

We presented a real-time noise reduction approach for low-delay and low-computation requirements.
Our overall delay is about \SI{8}{\ms}, consisting of about \SI{6}{\ms} filter bank analysis and synthesis, \SI{1}{\ms} future context and \SI{1}{\ms} processing.
While we are able to obtain similar results to other real-time approaches like RNNoise, we reduce the number of parameters significantly to only \SI{5}{\k} and number of \si{\FLOPS} to about \SI{10}{\M}.

\bibliographystyle{IEEEtran}
\bibliography{refs}

\end{document}